# A novel unsteady aerodynamic Reduced-Order Modeling method for transonic aeroelastic optimization

Ziyi Wang[a], Weiwei Zhang[a*], Xiaojing Wu[a1], Kongjin Chen[b]

[a] *School of Aeronautics, Northwestern Polytechnical University, China*
[b] *HONGDU Aviation Industry Group LTD, Aviation Industry Corporation of China, China*

**Abstract:**

In aircraft design, structural optimization and uncertainty quantification concerning transonic aeroelastic issues are computationally impractical, because the iterative process requires great number of aeroelastic analysis. Emerging Reduced-Order Model (ROM) method is convenient for transonic aeroelastic analysis. However, current ROMs cannot be reused during iteration, thus time cost is still way too large. This study proposed an improved ROM suitable for Arbitrary Mode Shapes (ROM-AMS), which is reusable regardless the variation of design variables. By adopting Principal Component Analysis, ROM-AMS method can significantly reduce the number of basis mode shapes and improve the accuracy of flutter analysis. In an optimization case, the weight of cropped delta wing is reduced by 28.46%, and the efficiency is 900 times higher than traditional ROM method, which demonstrates the feasibility of this method in iterative design process.

## 1 Introduction

In aerospace field, design optimization and uncertainty qualification (UQ) play an important role. Design optimization is carried out to obtain a subtle design with minimized weight or other objective features, meanwhile subject to necessary constraints such as safety. UQ is applied to quantitatively estimate the uncertainty, for instance design variable uncertainty, and impact of such uncertainty on the quantities of interest (Beran et al., 2017). In essence, design optimization and UQ are all iterative process, during which values of design variables are constantly updated and large number of tests are conducted.

Of interest here is structural optimization/UQ concerning transonic flutter constrains. Flutter is essentially the dynamic instability of aeroelastic systems. To avoid such disastrous failure in structure, flight speed should be lower than the critical flutter velocity (so called flutter boundary). In flutter analysis, the acquisition of unsteady aerodynamic response occupies the longest computation time, thus a rapid and reliable aerodynamic calculation method is of most importance for iterative process. At present, two branches are applied to calculate aerodynamic force in UQ/optimization: Traditional AeroElascity (TAE) and Computational AeroElasticity (CAE) (Beran et al., 2017). TAE predicts aerodynamic force rapidly using lower-fidelity aerodynamic tools. For example, Doublet Lattice Method (DLM) (Jutte et al., 2014; Werter and De Breuker, 2016; Wan et al., 2003; De Leon et al., 2012), strip theory (Georgiou et al., 2014; Weisshaar, 1981) and ONERA stall model (Stanford and Beran, 2013a) were applied to aeroelastic design optimization in subsonic regime, while piston theory was frequently used in supersonic and hypersonic regime for optimization and UQ based optimization (such as Reliability-Based Design Optimization, RBDO) (Stanford and Beran, 2013b, 2012).

On the other hand, many aircrafts, such as large commercial transport planes and fighters, need to cruise or combat in the transonic regime. Due to the flow nonlinearity, aircraft in transonic regime would undergo a dangerous reduction of flutter boundary, namely so called "transonic dip". However, flow nonlinearity such as shock wave and separated boundary layers cannot be correctly captured by TAE. As a compromise, Prandtl-Glauert correction factor is incorporated in TAE method (Mallik et al., 2013). But this method is still invalid when Mach number is approaching 1. With the development of Computational Fluid Dynamics (CFD), CAE becomes an appealing method in transonic regime and has been applied to design optimization concerning static-state aeroelastic problems (Barcelos and Maute, 2008). To obtain flutter velocity, traditional CAE analysis should be casted at many trial freestream velocities until a critical velocity is found. Such implement is computational impractical for iterative process and mainly used for post verification at present (Schuster et al., 2003; Yurkovich, 2003). To ease the problem, the flutter analysis was treated as eigenvalue problem by using Schur method, because flutter is actually Hopf-Bifurcation (Badcock and Woodgate, 2008). This implement has been utilized for UQ (Marques et al., 2010) and UQ based optimization (Marques and Badcock, 2012). In addition, (Chen et al., 2004; Stanford et al., 2015) proposed a novel field-panel method which is conductive to transonic aeroelastic optimization. First, CFD solver is launched to obtain a background steady flow. Linearization about the steady background flow is then carried out for a range of reduced frequencies and interpolated onto a flat-plate wing mesh with a field panel scheme (Stanford et al., 2015). Above process produces Aerodynamic Influence Coefficients (AIC) governing the relationship between pressure and downwash at a set of reduced frequencies. The AIC is independent of the variation of structural parameters, therefore it is reusable during structural optimization.

However, aforementioned Schur methods requires significant modifications to existing codes (Yao and Marques, 2017). Instead, unsteady aerodynamic Reduced-Order Model (Dowell and Hall, 2001; Lucia et al., 2004; Raveh, 2005) is applied in (Yao and Marques, 2017) to model nonlinear aerodynamic response. Existing ROM can be summarized in two branches: the Proper Orthogonal Decomposition (POD)/Dynamic Mode Decomposition (DMD) (Rowley et al., 2009; Schmid, 2010) method and the system identification method (such as Volterra series (Milanese and Marzocca, 2009) and Auto Regressive with eXogenous input model, ARX). At present, system identification based ROMs have been applied to aeroelastic problems extensively, for instance, flutter analysis at high angle of attack (Zhang and Ye, 2007), revelation of complex fluid mechanisms such as frequency lock-in in transonic buffeting flow (Gao et al., 2017b) and low Reynolds number vortex-induced vibrations (Zhang et al., 2015b), transonic Limit Cycle Oscillation (LCO) prediction (Mannarino and Mantegazza, 2014; Zhang et al., 2016, 2012), control low design for active flutter suppression (Chen et al., 2012), active control of transonic buffet flow (Gao et al., 2017a). In addition, a lot of novel nonlinear ROMs have been developed, for example: Kou proposed two layer ROM which can capture both linear and nonlinear aerodynamic responses (Kou and Zhang, 2017a), and Multi-kernel neural networks for nonlinear unsteady aerodynamic modeling (Kou and Zhang, 2017b). Maximilian Winter utilized neurofuzzy model to conduct unsteady aerodynamic computation in varying freestream conditions (Winter and Breitsamter, 2016).

Nevertheless, above method, including field-panel method and ROMs, are all fixed at Prescribed Mode Shapes (such ROM is referred as ROM-PMS) and not robust to mode shape variation in structure optimization. Existing ROMs can adapt to the change of modal mass or modal frequency (Song et al., 2011; Wang et al., 2008), but not adapt to the change of mode shapes. Marques demonstrated that neglecting the variation of mode shapes in structure-changeable cases will lead to misleading results (Marques et al., 2010). To solve the problem, Zhang et.al developed ROM suitable for arbitrary mode shapes (ROM-AMS) (Zhang et al., 2015a). First, a set of Radial Basis Functions (RBFs) are selected as basis mode shapes to linearly fit the physical mode shapes. Then, different from traditional ROM, ROM-AMS is constructed in basis mode coordinate. After being converted to real modal coordinate, ROM-AMS can replace CFD solver in aeroelastic analysis. For various structures with same aerodynamic shape and flow condition, ROM-AMS is reusable and robust to the variation of both modal frequency and mode shapes. Only one run of CFD process is required in structural optimization/UQ. Under the same framework, (Winter et al., 2017) proposed another ROM-AMS where Chebyshev polynomials are selected as basis mode shapes, which can capture the global features of physical mode shapes.

However, in (Winter et al., 2017; Zhang et al., 2015a), the number of required basis mode shapes is relatively large. Take ARX model as example, with same delay order, the number of parameters to be identified is square with the number of the basis modes. Consequently, excessive number of basis modes will significantly prolongs the time of training and modeling, especially in practical problems where the whole flight envelope in transonic Mach number interval is subject to the flutter boundary. However, less basis modes will enlarge the deviation of flutter prediction. So there is a tradeoff between the model simplicity and fitting accuracy. To make the tradeoff easier, PCA basis is adopted here as basis mode shapes. First, conduct parametric sampling and modal analysis on structures in design variable space. After this step, mode shapes belong to sample structures are selected as the input of PCA. Then, basis mode shapes are chosen from the PCA basis. Further analysis reveals that just small number of basis modes can reach desirable accuracy. Ultimately, ROM-AMS is applied in transonic aeroelastic design optimization for the first time. To verify the effectiveness of PCA basis, the variation of mode shapes is relatively large.

## 2 Numerical method

### 2.1 Basis mode shapes

In previous ROM-AMS, basis mode shapes are RBFs scattering in the wing plate. The shape of 2-dimensional RBF is shown in Fig1. After assigned correct coefficients, these RBFs can fit arbitrary mode shape in the wing plate. In (Zhang et al., 2015a), 18 RBFs were exploited, while the fitting accuracy is still less than perfect, which results in 9% deviation in flutter speed. If the number of basis modes continues to increase, the complexity and training time of the ROM will increase significantly. In light of this, the tradeoff seems difficult if RBFs are used as basis modes.

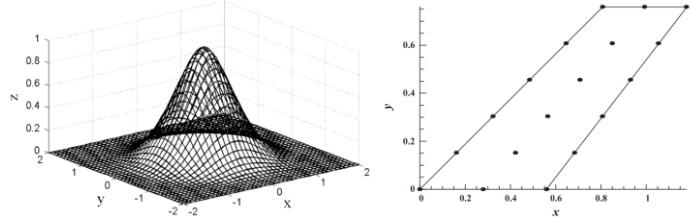

Fig1 Two dimension example and locations of RBF basis

To make the tradeoff easier, PCA is applied here to generate basis mode shapes. In some cases, PCA is also known as POD. First, parametric sampling is carried out in design variable space, and $l$ sample structures are obtained in this step. Then, after modal analysis, $n$ mode shapes $\Phi$ are acquired as the snapshots of PCA, where $n = l \times k$, and $k$ is the number of selected modes of each structure. Vector $\Phi$ represents the displacements at each node in the finite element model. As the number and location of nodes may vary in different structures, to form the snapshot matrix, mode shapes of sample structures must be expressed under a standard mesh, namely the normalization of mode shapes, which can be achieved by interpolation. Hence, the snapshot matrix can be formed by normalized mode shapes $\Phi_i$

$$P = [\Phi_1, \Phi_2, ..., \Phi_n]_{N \times n}$$

where $N$ is the number of nodes in standard mesh. The eigenvector matrix of $PP^T$ is

$$V = [\varphi_1, \varphi_2, ..., \varphi_N]$$

and the columns of $V$ are basis mode shapes. The corresponding coefficients $\gamma$ of an arbitrary mode shape $\Phi$ in design variable space can be obtained by solving

$$V\gamma = \Phi \tag{1}$$

To reduce the order of the model, $m$ eigenvectors $\varphi_i$ corresponding to the first $m$ largest eigenvalues are selected to approximate the real mode shape, $i = 1, 2...m$, namely

$$\Phi^* = \sum_{i=1}^{m} \gamma_i \varphi_i \tag{2}$$

Hence $N$-dimensional mode shape $\Phi$ can be reduced to $m$-dimensional vector $\gamma$. To determine a proper $m$, Modal Assurance Criteria is brought in and defined as below

$$\text{MAC} = \frac{\left(\Phi^{*T}\Phi\right)^2}{\left(\Phi^{*T}\Phi^*\right)\left(\Phi^T\Phi\right)} \geq \varepsilon \tag{3}$$

MAC=1 indicates a good agreement between simulated mode shape and real mode shapes. The minimum value of $m$ which satisfy (3) is chosen as the number of basis modes, where $\varepsilon$ is defined by user.

In some cases where $N$ is far greater than $n$, eigenvalue decomposition for $PP^T$ will be difficult, whereas $P^T P$ is much easier to deal with. Similarly, $m$ eigenvectors $v_i$ of $P^T P$ corresponding to the first $m$ largest eigenvalues are selected, which forms

$$V = [v_1, v_2, ..., v_m]$$

And the basis matrix is then formed as

$$F = PV\Lambda \tag{4}$$

where $\Lambda = diag(1/\sqrt{\lambda_j})$, and $\lambda_j$ is one of the $m$ largest eigenvalues, $j = 1, 2, ..., m$. The columns $\varphi_i$ in $F = [\varphi_1, \varphi_2, ..., \varphi_m]$ are the basis modes shapes. To determine a proper value of $m$, MAC is also applied.

After the basis mode shapes being obtained, the $j$ th mode shape of arbitrary structure in design variable space can be expressed as

$$\boldsymbol{\Phi}_j = \sum_{i=1}^{m} \gamma_{ij} \boldsymbol{\varphi}_i \tag{5}$$

where $\gamma_{ij}$ can be obtained by solving linear equations.

In summary, there are two reasons accounting for why PCA method can simulate mode shapes with fewer basis modes. 1) In contrast to RBF and Chebyshev polynomials, PCA basis are customized for specific problem. 2) Although sometimes the mode shapes vary greatly in iterative process, they still obey some basic patterns. For example, in section 3.1, the second order mode shapes of different structures perform a torsional commonness. Such basic patterns can be easily identified by PCA.

## 2.2 ROM-AMS

First, $m$ excitation signals $u_i(t)$ are designed for $m$ basis mode shapes $\boldsymbol{\varphi}_i, i=1,2,...,m$, then CFD solver is launched to calculate the corresponding aerodynamic responses. The actual displacement of wing boundary mesh during calculation is expressed as

$$\boldsymbol{d}(t) = \sum_{i=1}^{m} u_i(t) \boldsymbol{\varphi}_i \tag{6}$$

Then the corresponding response, namely Generalized Aerodynamic Coefficients (GAC) are calculated as below

$$\begin{aligned} f_1 &= \iint P\boldsymbol{\varphi}_1 ds \Big/ q \\ f_2 &= \iint P\boldsymbol{\varphi}_2 ds \Big/ q \\ &\vdots \\ f_m &= \iint P\boldsymbol{\varphi}_m ds \Big/ q \end{aligned} \tag{7}$$

Hence the excitation-response data set has been obtained. For the discrete-time multi-input multi-output system, the ARX model is selected to construct the unsteady aerodynamic ROM based on basis mode shapes

$$\boldsymbol{f}(t) = \sum_{i=1}^{na} \boldsymbol{A}_i \boldsymbol{f}(t-i) + \sum_{i=0}^{nb-1} \boldsymbol{B}_i \boldsymbol{u}(t-i) \tag{8}$$

where $t$ denotes the time step, $\boldsymbol{u}$ stands for the input, namely the generalized displacements of basis modes, and $\boldsymbol{f}$ represents the output, namely the GAC of basis modes. $\boldsymbol{A}_i$ and $\boldsymbol{B}_i$ are the constant coefficient matrices to be estimated, $na$ and $nb$ are delay orders which is determined by user. Once the training data is obtained, $\boldsymbol{A}_i$ and $\boldsymbol{B}_i$ can be calculated by using the least square method, hence the unsteady aerodynamic ROM on basis mode shapes is constructed. As the scale of $\boldsymbol{A}_i$ and $\boldsymbol{B}_i$ grows quadratically with increasing $m$, the training time of ARX model should also quadratically increases. In light of this, the number of basis mode shapes should be reduced as much as possible.

However, it should be emphasized that current ROM is based on basis mode shapes, and cannot be used in real modal coordinate. Therefore generalized displacements and GAC should be transformed between real modal coordinate and basis modal coordinate

$$u_1 = \sum_{j=1}^{k} \xi_j \gamma_{1j}$$

$$u_2 = \sum_{j=1}^{k} \xi_j \gamma_{2j} \qquad (9)$$

$$\vdots$$

$$u_m = \sum_{j=1}^{k} \xi_j \gamma_{mj}$$

$$F_1 = \iint P\Phi_1 \mathrm{d}s \Big/ q = \iint P \sum_{i=1}^{m} \gamma_{i1}\varphi_i \mathrm{d}s \Big/ q$$

$$= \sum_{i=1}^{m} \gamma_{i1} \iint P\varphi_i \mathrm{d}s \Big/ q = \sum_{i=1}^{m} \gamma_{i1} f_i$$

$$\vdots \qquad (10)$$

$$F_k = \iint P\Phi_k \mathrm{d}s \Big/ q = \iint P \sum_{i=1}^{m} \gamma_{ik}\varphi_i \mathrm{d}s \Big/ q$$

$$= \sum_{i=1}^{m} \gamma_{ik} \iint P\varphi_i \mathrm{d}s \Big/ q = \sum_{i=1}^{m} \gamma_{ik} f_i$$

Equation (9) transforms real generalized displacements to basis modal coordinate, and (10) transforms basis GAC back to real modal coordinate. Only by this way can ROM-AMS be coupled in real modal coordinate.

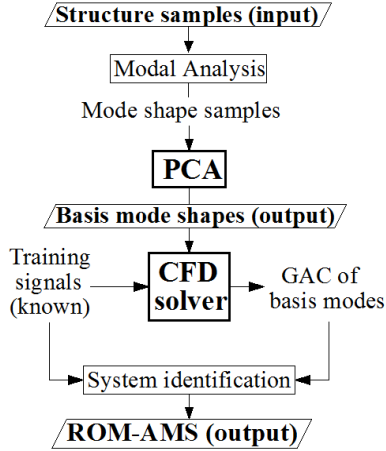

Fig2 Construction of ROM-AMS on basis modes

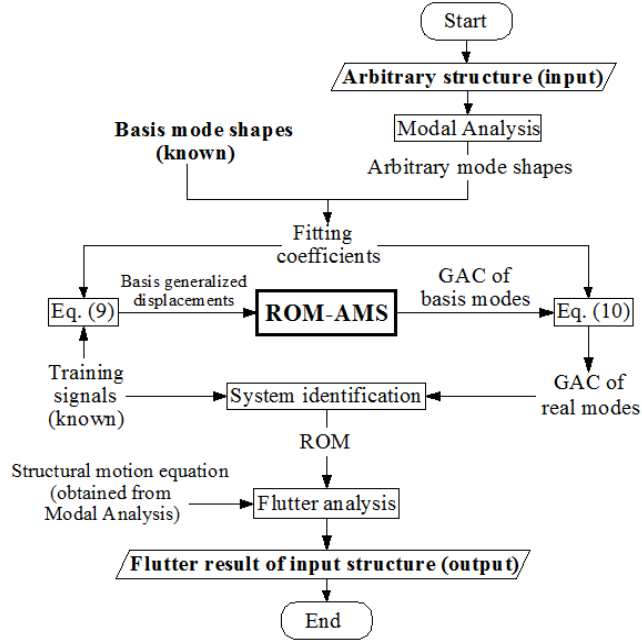

Fig3 Flowchart of ROM-AMS and flutter analysis

Given a training signal in real mode coordinate for arbitrary structure, ROM-AMS can calculate responses without another run of CFD. Thus the excitation-response data set are obtained to train classical ROM in real modal coordinate, which can be written in continuous-time state-space form

$$\begin{cases} \dot{x}_a(t) = A_a \cdot x_a(t) + B_a \cdot \xi(t) \\ F(t) = C_a \cdot x_a(t) + D_a \cdot \xi(t) \end{cases} \quad (11)$$

and the structural motion equation under modal coordinate is written as

$$M \cdot \ddot{\xi} + G \cdot \dot{\xi} + K \cdot \xi = q \cdot F \quad (12)$$

where structural characteristics $M$, $G$ and $K$ should be determined through modal analysis. Convert (12) to continuous-time state-space form

$$\begin{cases} \dot{x}_s(t) = A_s \cdot x_s(t) + q \cdot B_s \cdot F(t) \\ \xi(t) = C_s \cdot x_s(t) + q \cdot D_s \cdot F(t) \end{cases} \quad (13)$$

Coupling (11) and (13), the aeroelastic equation can be written as

$$\dot{x} = A \cdot x = \begin{bmatrix} A_s + q \cdot B_s \cdot D_a \cdot C_s & q \cdot B_s \cdot C_a \\ B_a \cdot C_s & A_a \end{bmatrix} \cdot x \quad (14)$$

Hence the stability analysis of aeroelastic system can be transformed to the eigenvalue problem of $A$. To analysis the flutter characteristics, $vg$ and $v\omega$ plot are obtained by solving the eigenvalue of $A$ under different dynamic pressure. The detailed derivation can be found in (Zhang et al., 2015b).

## 2.3 Differential Evolution algorithm

Differential evolution algorithm (DE) (Storn and Price, 1997) is applied in present work to optimize a wing structure. Comparing with gradient algorithm, DE algorithm can capture a global optimal solution. First, the collection of all structural parameters is considered as an individual, namely $x = [x_1, x_2, ... x_D]$, where $x_i$ represents the $i$th parameter, namely the "gene" of individual. $D$ is the total gene number of an individual. Then, $NP$ individuals are generated by randomly changing the structural parameters in the optimization space to construct the initial population $XG_0$.

Undergoing 3 steps listed below, the next generation of population is obtained:

(1) Mutation: In the $g$th generation $XG_g$ ($XG_0$ for the first iteration), mutation is conducted as

$$v_i(g+1) = x_{r1}(g) + F \bullet (x_{r2}(g) - x_{r3}(g)) \quad (15)$$

where $v_i(g+1)$ means the $i$th mutant individual, $r1, r2, r3 \in \{1, 2, 3, ..., Np\}$ are 3 random integers different from $i$, $F \in [0, 2]$ is scale factor. Once $v_i(g+1)$ is beyond the boundary, this individual should be created again until boundary constraint is satisfied. The collection of $v_i(g+1)$ forms the first trial population $XG_{next1}$;

(2) Crossover: $x_i(g) = (x_{1i,g}, x_{2i,g}, ..., x_{Di,g})$ and $v_i(g+1) = (v_{1i,g+1}, v_{2i,g+1}, ..., v_{Di,g+1})$ are picked out from $XG_g$ and $XG_{next1}$ to generate an new individual $u_i(g+1) = (u_{1i,g+1}, u_{2i,g+1}, ..., u_{Di,g+1})$, $i = 1, 2, ..., Np$. The $j$th gene of $u_i(g+1)$ are determined by the law

*if* $rand \leq CR$ *or* $randj = j$
$u_{ji,g+1} = v_{ji,g+1}$
*else*
$u_{ji,g+1} = x_{ji,g}$

where $rand$ is a random number $\in [0,1]$, and $randj$ is a random integer $\in [0, D]$. $CR$ is the crossover constant $\in [0,1]$ which is defined by user. The collection of all the $u_i(g+1)$ forms the second trial population $XG_{next2}$;

(3) Selection: $x_i(g)$ and $u_i(g+1)$ are selected from $XG_g$ and $XG_{next2}$ respectively and then compared with each other. The individual whose objective function is lower is selected into next generation. When the comparison is accomplished at all index, population of next generation $XG_{g+1}$ is updated and ready for next iteration;

Iteration of the 3 steps should be kept until convergence, as is shown in Fig4.

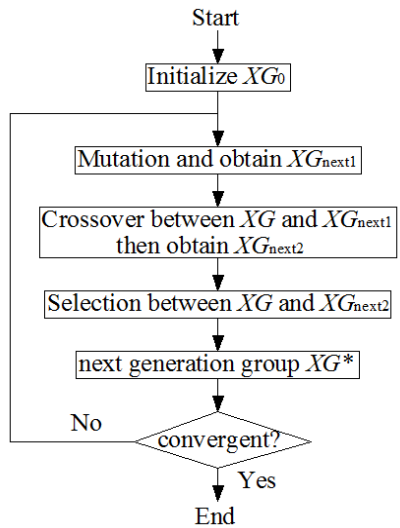

Fig4 Flowchart of Differential Evolution Algorithm

## 3 Numerical Example

The prototype of the model is cropped delta wing (Schairer and Hand, 1999) which is a typical experimental and numerical case for transonic LCO analysis (Attar and Gordnier, 2006; Chen et al., 2014; Gordnier, 2002; GORDNIER and MELVILLE, 2001; Peng and Han, 2011). The aerodynamic features are governed by leading-edge vortex and shock wave (GORDNIER and MELVILLE, 2001). The structural weight is about 0.24$kg$. As is shown in Fig5, with a beam and lumped mass installed, the original flat-plate wing is altered to parameter-changeable model. Because ROM-AMS is not robust to the variation of aerodynamic shape, the aerodynamic shape is assumed unchangeable during optimization. The framework of optimization is seeking a set of optimal structural parameters to define a lightest structure, meanwhile satisfying the flight envelop which is subject to flutter boundary. To verify the effectiveness of ROM-AMS, 7 design variables are deliberately chosen so that the modal frequency and mode shapes are sensitive to the variation of design variables. Among these variables: $x_1$ and $x_2$ define the position of both end of the beam; $x_3$ defines the spanwise location of lumped mass which is installed at the leading edge, and $x_4$ is the weight of lumped mass; $x_5$ and $x_6$ define the sectional shape of the beam, and $x_7$ is the thickness of wing plate. The detailed definition and the variable space are shown in Fig 5(b) and Table1. To clearly illustrate the variables, Fig 5(b) magnifies the beam and wing plate thickness in an exaggerated way. Table2 compares experimental and computational modal frequencies, which verifies the accuracy of Nastran solver.

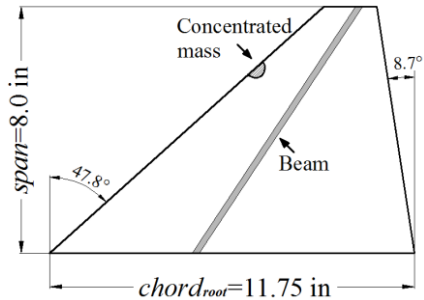

(a) Layout of the parameter changeable wing

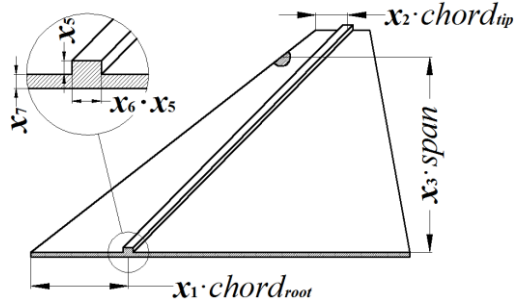

(b) Illustration of design variables

Fig 5 The parameter changeable wing to be optimized

Table1 Definitions of design variables

| Parameter | Definition(unit) | Lower bound | Upper bound |
|---|---|---|---|
| $x_1$ | Root end position of the beam (ratio of $chord_{root}$) | 15% | 70% |
| $x_2$ | Tip end position of the beam(ratio of $chord_{tip}$) | 15% | 70% |
| $x_3$ | Position of the lumped mass(ratio of $span$) | 60% | 85% |
| $x_4$ | Weight of lumped mass($kg$) | 0 | 0.105 |
| $x_5$ | Sectional height of the beam($mm$) | 0.5 | 2 |
| $x_6$ | Sectional width of the beam(ratio of $x_5$) | 50% | 200% |
| $x_7$ | Thickness of plate(ratio of thickness in(Schairer and Hand, 1999)) | 50% | 100% |

Table 2 The modal frequency of the original flat plate wing (Hz)

|  | Mode1 | Mode2 | Mode3 |
|---|---|---|---|
| Nastran | 26.95 | 88.72 | 133.16 |
| Experiment(Schairer and Hand, 1999) | 26.7 | 88.2 | 131.8 |

### 3.1 Accuracy test of Reduced Order Model

The accuracy of ROM-PMS will be examined first in this section. Original flat plate wing is applied here for the comparison to experimental result. To construct ROM, signals in Fig 6 are utilized as excitation of the first 4 structural modals of flat plate wing. The freestream Mach number is 0.87. (GORDNIER and MELVILLE, 2001) indicates that the viscous influence is small for lower amplitude of structural deflection (just like the training signals), hence Mach number is adequate to define an identical flow as the experiment, and atmospheric parameter at 10km altitude is given as the free stream condition.

To verify the grid resolution, two sets of grids are generated: baseline and refined grid. Both grids are based on the surface mesh shown in Fig 7, which contains 10976 cells in one side. The resolution of surface mesh is slightly higher than its counterpart in (Peng and Han, 2011). Different from baseline grid, refined grid contains 30 layer of prisms in the vicinity of the surface mesh. Baseline grid contains about 0.4 million cells, while refined grid contains 1.2 million. Results of ROM-PMS method as well as experiment are listed in Table 3, which indicate that ROM-PMS method using baseline grid reaches enough accuracy and can be regarded as standard solution. The accuracy of ROM-PMS was also proven in (Zhang et al., 2015a). In addition, the flutter dynamic pressure of linearized aerodynamic method is listed in Table3 marked by p-k method (Schairer and Hand, 1999). Because of the transonic feature, linearized method cannot obtain accurate result. To keep consistent with experiment, psi (1.0 psi=6895.0 Pa) is applied in Table 3 as the unit of dynamic pressure.

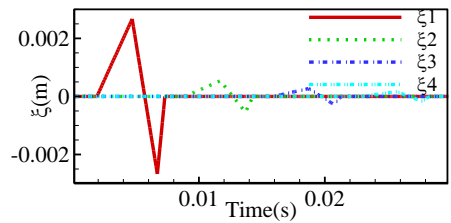

Fig 6 Modal excitation signals of original flat plate wing

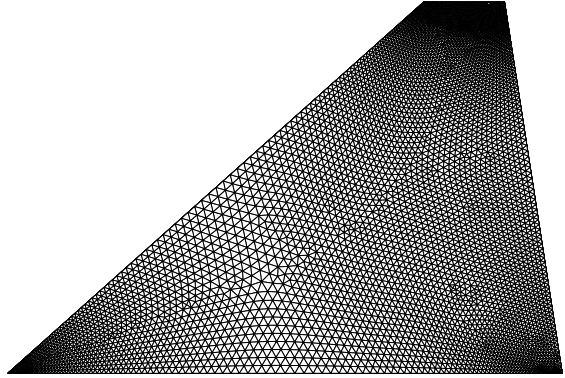

Fig 7 Surface mesh of aerodynamic model

**Table 3 Results of different flutter analysis method of flat plate wing (Ma=0.87)**

| Method | Flutter dynamic pressure (psi) |
|---|---|
| ROM-PMS(baseline grid) | 2.488 |
| ROM-PMS(refined grid) | 2.458 |
| p-k (Schairer and Hand, 1999) | 2.750 |
| Experiment (GORDNIER and MELVILLE, 2001; Schairer and Hand, 1999) | 2.400 |

Next, to verify the accuracy of ROM-AMS and its capability for large variation of mode shapes, two typical structures are chosen from design space. Their corresponding parameters are $var_1$ and $var_2$. The mode shapes of two structures are compared in Fig8. Clearly, the modal frequency and mode shapes vary significantly with the change of structural parameter, and higher order modal gives more complex shape. To obtain basis mode shapes, Latin hypercube sampling is executed and 25 samples are created. Fig 9 illustrates 8 PCA basis calculated from those samples.

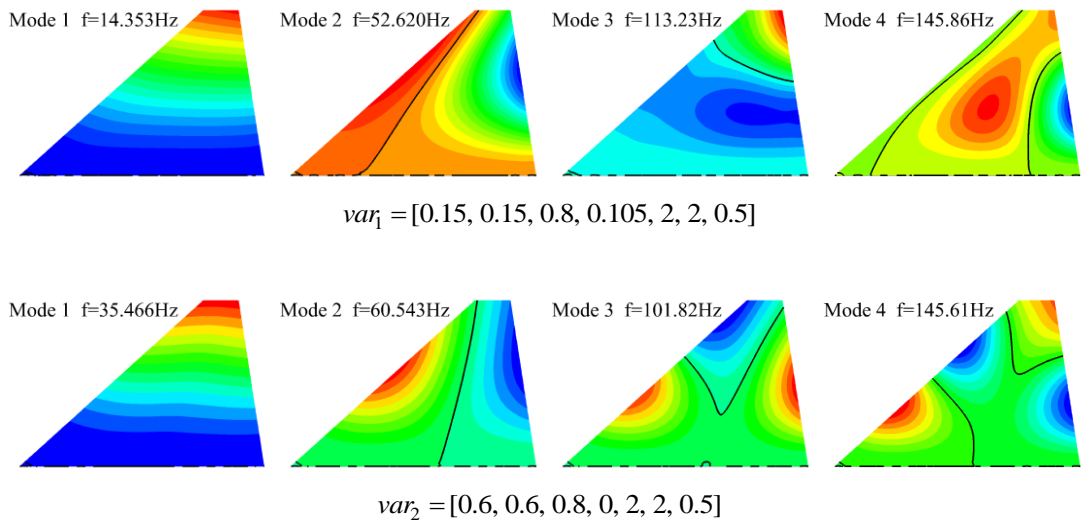

Fig8 The mode shapes of structures with $var_1$ and $var_2$

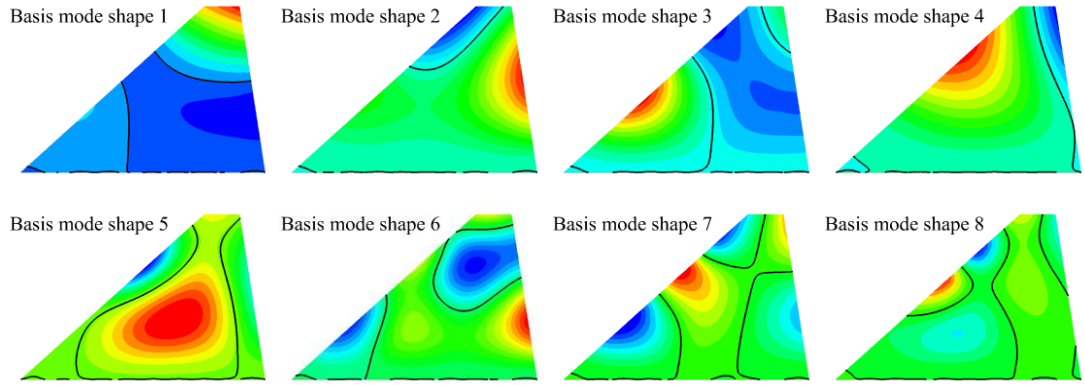

Fig9 First 8$^{th}$ PCA basis (basis mode shapes)

To characterize the fitting accuracy, MAC values of above structures are listed in Table4, and Fig10 compares the original and fitted mode shapes. The lower order mode shapes are easier to fit thus no need to be shown in Fig10. Fig11 indicates that more basis modes give higher fitting accuracy. Take $var_1$ as an example, when the basis number is lower than 7, the fitting accuracy is not adequate, especially for higher order modals which are difficult to fit. As the number of selected basis increasing, MACs of all modes are approaching to 1. 8 basis modes are used in this article, thus the fitting accuracy is adequate.

**Table4 The MAC values of structures with $var_1$ and $var_2$**

| Sample number | Mode1 | Mode2 | Mode3 | Mode4 |
|---|---|---|---|---|
| $var_1$ | ≈ 1 | 0.9996 | 0.9956 | 0.9886 |
| $var_2$ | ≈ 1 | 0.9996 | 0.9986 | 0.9872 |

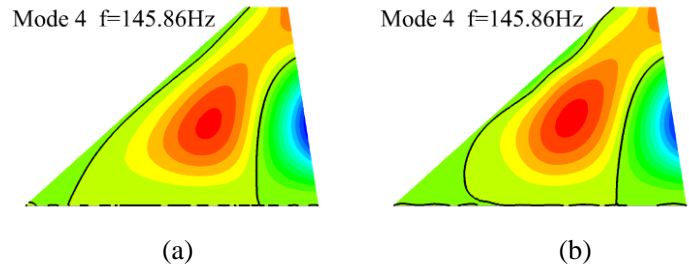

(a)          (b)

Fig10_1 Comparison between real(a) and simulated(b) mode shapes of $var_1$ structure

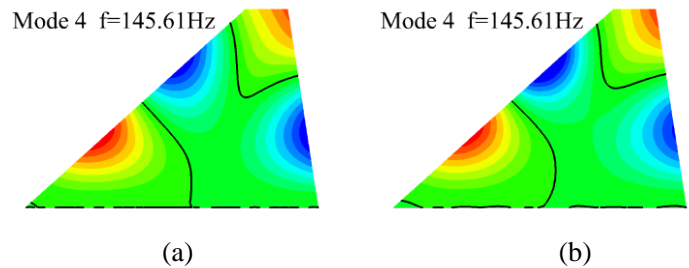

(a)          (b)

Fig10_2 Comparison between real(a) and simulated(b) mode shapes of $var_2$ structure

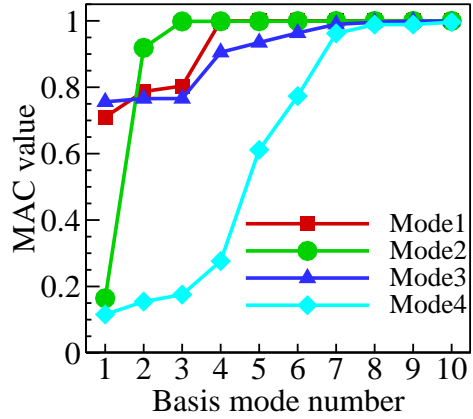

Fig11 The relationship between basis mode number and MAC value (structure with *var*$_1$)

Under same flow condition, ROM-AMS is trained by signals in Fig12. Model orders *na* and *nb* are all equal to 8. Fig13 verifies the accuracy of ROM-AMS by comparing the direct aerodynamic responses and ROM-AMS predicted responses. Flutter results of two structures are listed in Table5.

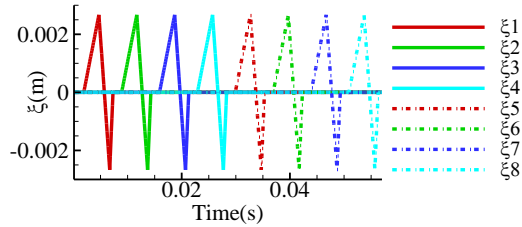

Fig12 Modal excitations of basis mode shapes

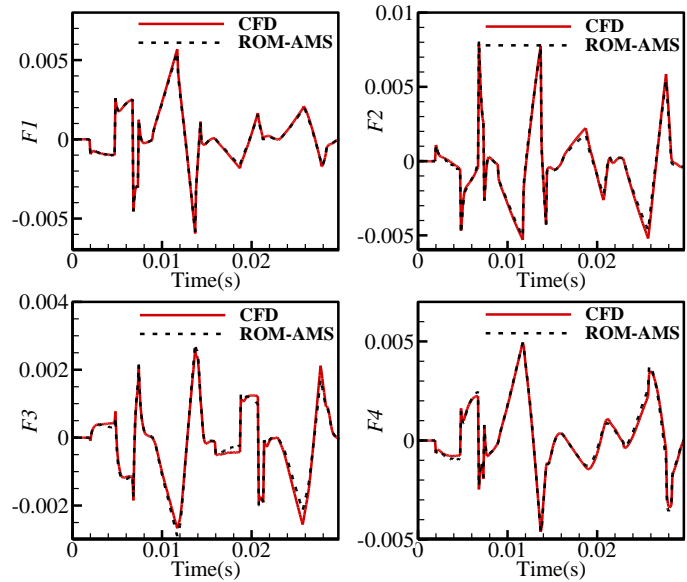

Fig13 Aerodynamic responses of the modal excitations in Fig9 (structure with *var*$_1$)

**Table 5 The comparison of flutter results between ROM-AMS and ROM-PMS (Ma=0.87)**

|      | Critical Flutter Speed(m/s) | | | Flutter frequency(rad/s) | | |
| --- | --- | --- | --- | --- | --- | --- |
|      | ROM-AMS | ROM-PMS | Error(%) | ROM-AMS | ROM-PMS | Error(%) |
| Var1 | 219.53 | 214.68 | 2.26 | 561.60 | 550.05 | 2.20 |
| Var2 | 148.56 | 150.47 | -1.27 | 252.95 | 254.03 | -0.43 |

## 3.2 Optimization case

In transonic regime, it is flutter boundary that constrains flight envelop. For example, to design a flight envelop for original flat-plate wing, critical dynamic pressure under 5 Mach numbers is calculated at first. The 5 Ma numbers are {0.79, 0.83, 0.87, 0.91, 0.95}. The maximum safe dynamic pressure is assumed 75.6% of critical dynamic pressure according to the typical criteria that the critical flutter speed should be at least 15% higher than the maximum allowable speed. Then, the corresponding safe altitude, namely the flight envelop, of original model under 5 Mach numbers is plotted in Fig 14, which divides airspace into safe zoom and dangerous zoom.

In the following optimization, flight envelop of original flat-plate wing is given as the benchmark for constraint. Only those structures whose flight envelop is lower than benchmark can be identified as feasible structure. For example, the envelop marked with "safe envelop" is lower than benchmark envelop, which means this envelop has a broader safe zoom, and structure having such envelop is safer than original wing. On contrary, the flight envelop across the benchmark boundary should be marked with "dangerous" because the safe zoom is narrower when $Ma<0.9$. In some other cases, flight envelop of a structure is much lower than the benchmark, which implies structural redundancy and the weight can be further reduced. In summary, the structure should be deliberately designed so that satisfying the constraint without redundant weight.

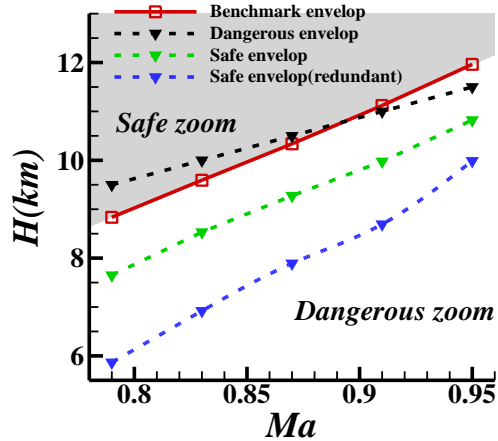

Fig 14 The illustration of benchmark flight envelop

Parameters of DE are set as: $F$=0.55, $Np$=30, $CR$=0.7. 7 parameters are defined in this model, therefore $D$=7. The original problem is transformed into unconstrained optimization problem by penalty function method, namely

$$Object = f(\boldsymbol{x}) + \mu \bullet C(\boldsymbol{x})$$

where $f$ is structural mass, $\mu$ is the penalty factor which is usually a large number. $C$ is penalty function: when the constraints are satisfied, $C$=0，otherwise $C$=1. In this work, constraints are: 1) flight envelop is lower than benchmark; 2) no single degree of freedom flutter (SDOFF) arises; 3) MAC values of all the modals should not be lower than 0.98.

Fig15 is the history of objective function. As $C(\boldsymbol{x})$ maintains 0 during iteration, objective function is actually the total mass of structure. In contrast to original flat-plate model with mass of 0.24$kg$, weight of optimal structure is 0.1717kg, which is reduced by 28.46%. Fig16 is the history of design variables (nondimensionalized by optimal values), and the optimal design variables are listed in Table 6.

Table 6 Optimal design variables

| | $x_1$ | $x_2$ | $x_3$ | $x_4(kg)$ | $x_5(mm)$ | $x_6$ | $x_7$ |
|---|---|---|---|---|---|---|---|
| Optimal value | 0.1535 | 0.6135 | 0.8247 | 0.0203 | 1.966 | 0.8787 | 0.5972 |

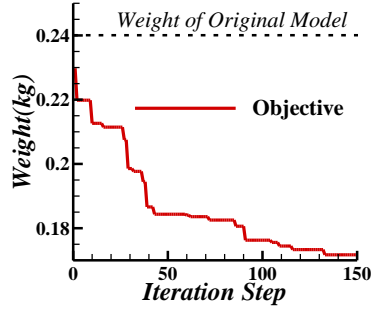

Fig15 The convergence history of structural weight (objective)

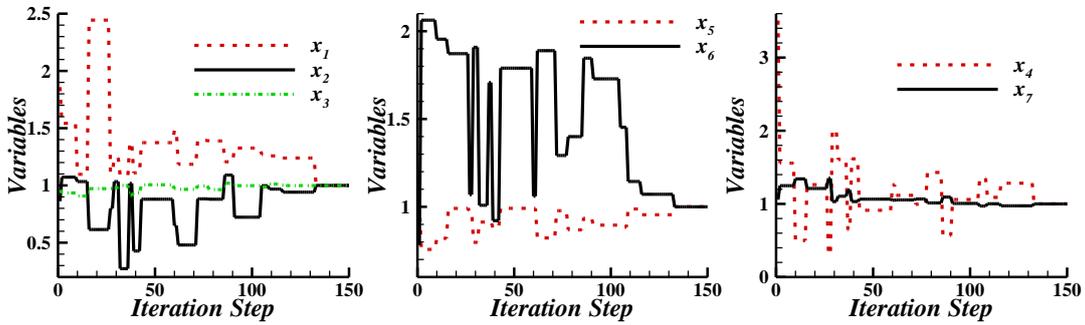

Fig 16 The convergence history of dimensionless design variables

The population distribution of initial generation ($G=0$), 50th generation ($G=50$) and the final generation ($G=150$) is demonstrated in Fig17. The horizontal axis represents the sample number, and vertical axis means the actual value of the genes (namely design variables). It can be found that in the early stages of evolution, genetic values scatter in design space. However, with the evolution pushing forward, the population gradually converges, and finally huddle in a narrow interval (except $x_2$ and $x_6$), which indicates the algorithm is nearing termination.

However, not all variables can eventually converge. For example, the populations of $x_2$ and $x_6$ still scatter when $G=150$ as is shown in Fig17. The reason is: these variables don't have substantial impact on objectives and constrains, which means they can be arbitrary value. These variables cannot evolve since there is no different between 'good gene' and 'bad gene'. It is difficult to know in advance which variables will have an important impact in preliminary design stage, they must be found by observing their variance. In subsequent design process, these untypical variables should be excluded so that the optimization will be clearly-oriented.

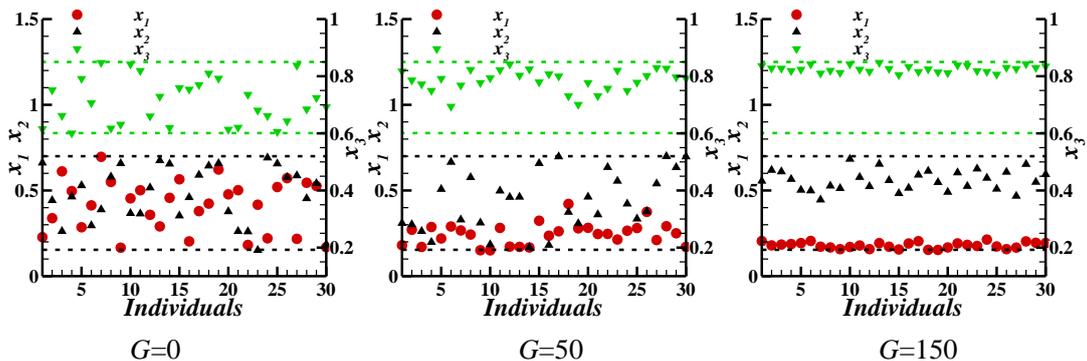

| $G=0$ | $G=50$ | $G=150$ |

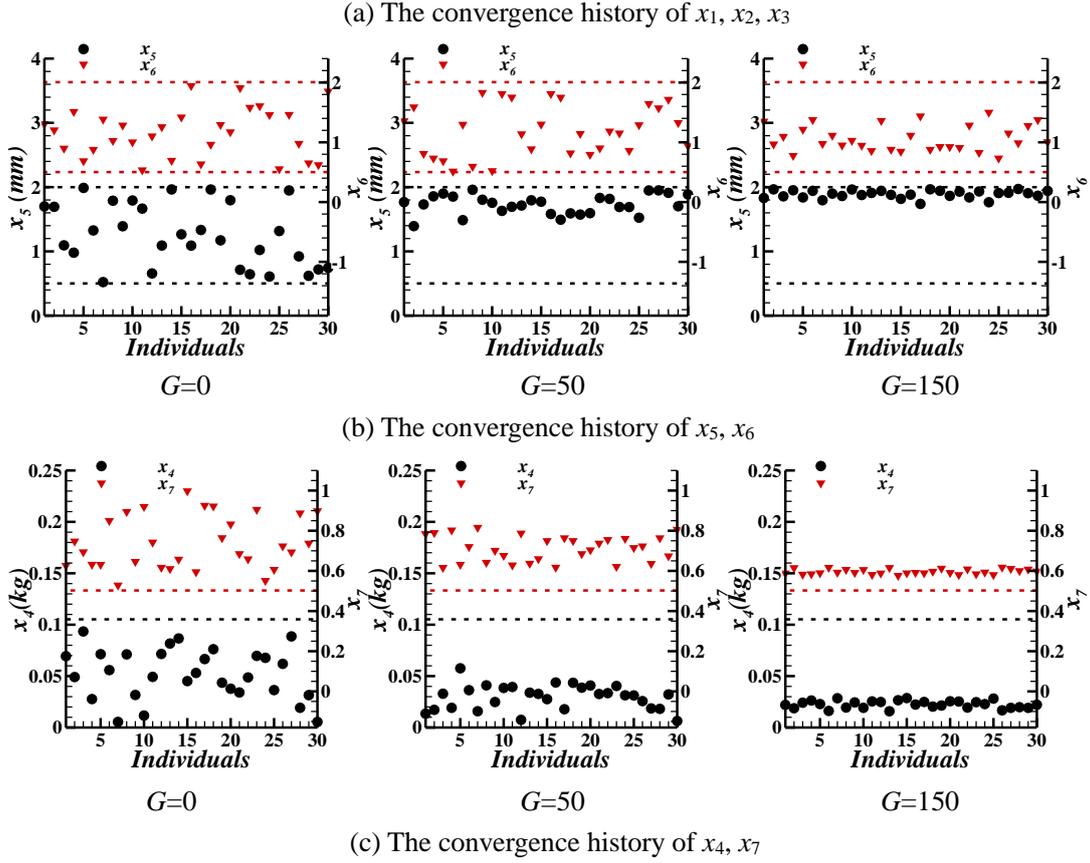

(a) The convergence history of $x_1$, $x_2$, $x_3$

(b) The convergence history of $x_5$, $x_6$

(c) The convergence history of $x_4$, $x_7$

Fig17 The convergence history of design variable population

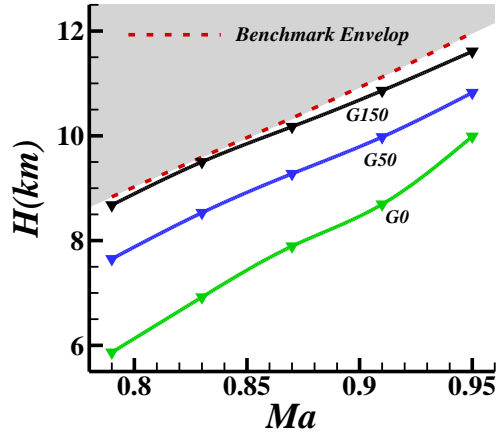

Fig 18 The convergence history of flutter altitude boundary

Fig18 demonstrates the history of flight envelop in different generation. $G_0$ is the flight envelop of the optimal individual in initial generation, and $G_{50}$, $G_{150}$ stand for the 50th and the 150th (final) generation respectively. It can be found that the flight envelop gradually boosts and attaches to the benchmark boundary, and the space for further optimization is decreasing. In $G_{150}$, the flight envelop of optimal structure is almost equal to the benchmark envelop, which indicates the optimization is ready to terminate. The flutter results of optimal structure is verified as is shown in Table7 and Table8.

4530 different structures were analyzed during optimization, and the time cost on PC with 8 cores of 3.6GHz and 8G RAM is 57h33min. Table9 compares the time cost for one run of structural analysis through ROM-PMS and ROM-AMS. For a single structure, calculation of aerodynamic responses by CFD takes up the majority of cost time. It is the repetition of flow solving that obstructs ROM-PMS from practical optimization, because the computational cost will be 5.91 years for 4530 structures in above example, in contrast to 57h33min for ROM-AMS.

Table7 MAC values of optimal structure

|  | Mode1 | Mode2 | Mode3 | Mode4 |
| --- | --- | --- | --- | --- |
| MAC | ≈ 1 | 0.9999 | 0.9994 | 0.9905 |

Table8 Calculation accuracy of optimal structure

| Ma | Critical Flutter Speed(m/s) | | | Flutter frequency(rad/s) | | |
| --- | --- | --- | --- | --- | --- | --- |
|  | ROM-AMS | ROM-PMS | Error(%) | ROM-AMS | ROM-PMS | Error(%) |
| 0.79 | 302.08 | 312.21 | -3.24 | 228.13 | 231.93 | -1.64 |
| 0.83 | 298.42 | 307.06 | -2.81 | 220.73 | 224.14 | -1.52 |
| 0.87 | 297.02 | 299.45 | -0.81 | 214.94 | 213.86 | 0.51 |
| 0.91 | 294.36 | 288.32 | 2.09 | 207.90 | 200.87 | 3.50 |
| 0.95 | 289.63 | 277.62 | 4.33 | 198.87 | 188.74 | 5.37 |

Table 9 Time cost comparison between different ROMs (in minute except last row)

| Process | ROM-PMS | | ROM-AMS | |
| --- | --- | --- | --- | --- |
|  | single structure | 4530 structures | single structure | 4530 structures |
| PCA | 0 | 0 | 13 | 13 |
| Modal Analysis | 0.5 | 0.5•4530 | 0.5 | 0.5•4530 |
| CFD | 685 | **685•4530** | 1350 | **1350** |
| Flutter Analysis | 0.15 | 0.15•4530 | 0.42 | 0.42•4530 |
| In total | 11.43h | 5.91 years | 22.73h | 57h33min |

Lastly, the conclusion in (Marques et al., 2010) is verified as below, which revealed the fact that the ignorance of variation of mode shape in optimization will lead to erroneous results. Assuming in above optimization, the mode shapes are fixed and just the modal frequency is altered during optimization. The fixed mode shapes are taken from the original flat plate wing. The flutter analysis of optimal structure with fixed mode shape at Ma=0.87 is conducted as below, whereas modal frequency keeps real value. Real mode shapes and assumed fixed mode shapes are compared in Fig19. The outcome reveals that the flutter speed will be 207.4m/s once the mode shapes are fixed, which is well below the standard solution 297.02m/s (in table8). The reason is apparent: the variation of nodal line in Mode 2 is not considered without ROM-AMS. Fig20 demonstrates the significant deviation of flutter speed in vgω plot. Therefore, to conduct authentic aeroelastic optimization, variation of mode shapes must be considered.

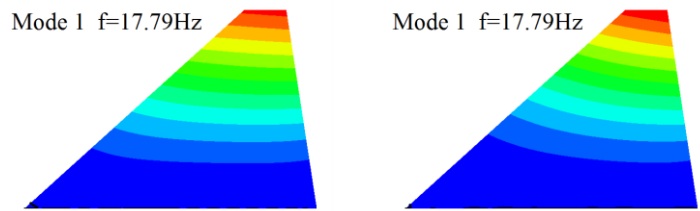

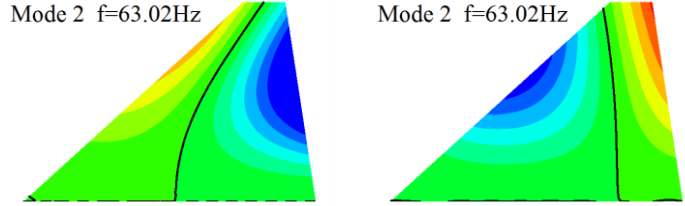

(a) real frequency and mode shape    (b) real frequency and fixed mode shape

Fig19 The comparison between real and fixed mode shapes of optimal structure

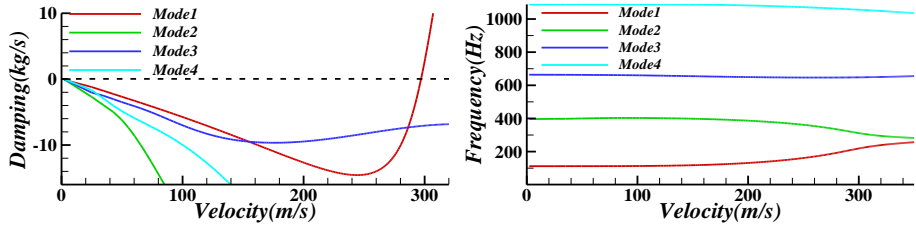

(a)

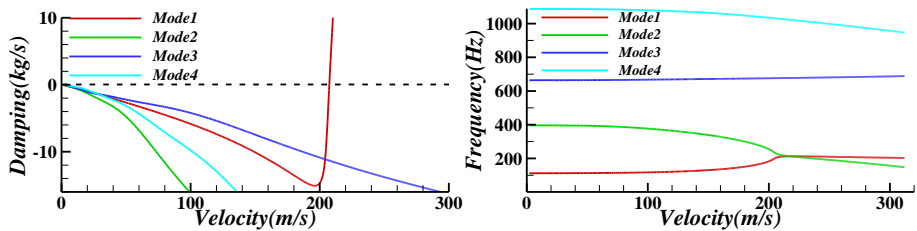

(b)

Fig20 The vgω plot of optimal structure at Ma=0.87
(a)ROM-AMS; (b)Analysis with fixed mode shape

## 4 Conclusion

This article proposed an unsteady aerodynamic reduced-order modeling method which is suitable for arbitrary structure (arbitrary mode shape) in design variable space, and applies this method to structural design optimization subject to transonic flutter constraint. The ROM is based on basis mode shapes by which the real mode shapes of parameter-changeable structure can be fitted, therefore it can be reused even if values of design variables are altered. In this work, Principal Component Analysis of structure samples in design variable space is first conducted, then PCA basis are selected as basis mode shapes. The accuracy tests of two representative structures demonstrate that just small number of PCA basis are required for accurate modeling, in comparison of other types of basis modes. The flutter predictions of those structures agree well with the result of traditional ROM with prescribed mode shapes. In optimization case where 4530 structures are analyzed, structural weight is reduced by 28.46%, while the efficiency is 900 times faster than traditional modeling method. The accuracy test of optimum structure manifests same order precision with traditional modeling method in transonic regime. Lastly, Marques's conclusion is verified, which demonstrates that the variation of mode shapes cannot be neglected in iterative process. At present, ROM-AMS is constructed on the same aerodynamic shape and flow parameters (such as Mach number and static angle of attack). For example, to conduct aeroelastic analysis in a flight envelop, several ROMs should be built at representative Mach numbers. In this case, PCA based approach has greater advantage because less basis number requires less training time.

## Acknowledgments

This work was supported by the National Science Fund for Excellent Young Scholars [11622220]; the 111 project [B17037]; and the ATCFD project [2015-F-016].

## Appendix:

In this appendix, the accuracy of ROM-AMS is verified by Euler solver. The Critical Flutter Velocity (CFV) intervals calculated by CFD solver under each Mach number are listed as below, where the percentage in Table A.1 denotes the ratio of CFD result with respect to ROM-AMS result, namely

$$ratio = \frac{CFV_{CFD}}{CFV_{ROM-AMS}} \tag{A.1}$$

**Table A.1 Flutter intervals calculated by CFD**

| Ma | Lower bound | Higher bound |
|---|---|---|
| 0.79 | 100% | 103% |
| 0.83 | 95% | 100% |
| 0.87 | 95% | 100% |
| 0.91 | 90% | 95% |
| 0.95 | 87% | 90% |

The temporal responses calculated by CFD solver under different Mach numbers are demonstrated as below.

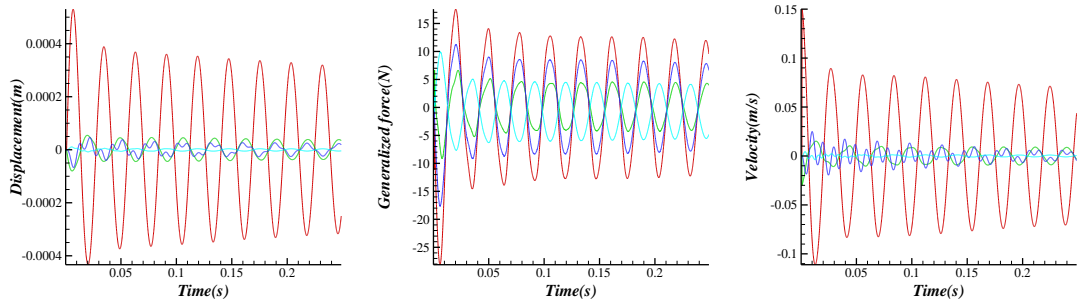

Fig A.1 $Ma$=0.79, $V$=100% $CFV_{ROM\text{-}AMS}$

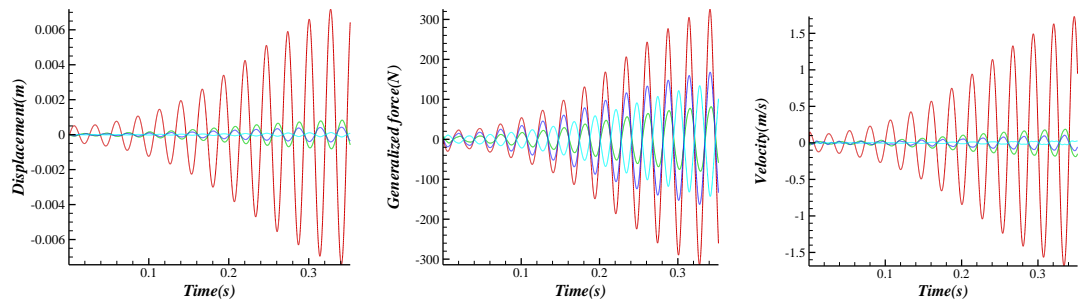

Fig A.2 $Ma$=0.79, $V$=103%$CFV_{ROM\text{-}AMS}$

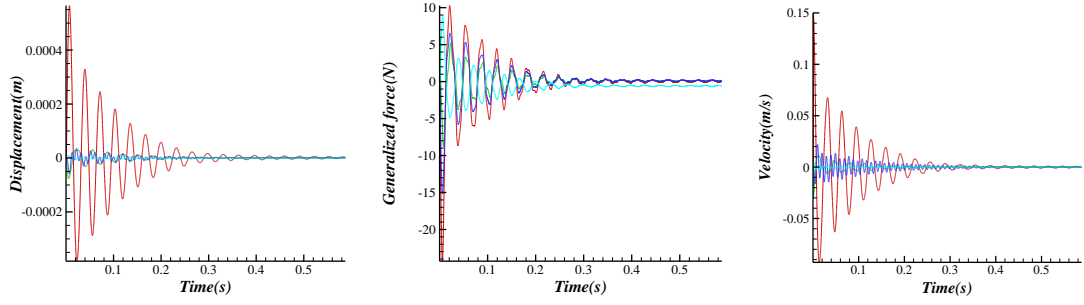

Fig A.3 $Ma$=0.83, $V$=95%$CFV_{ROM\text{-}AMS}$

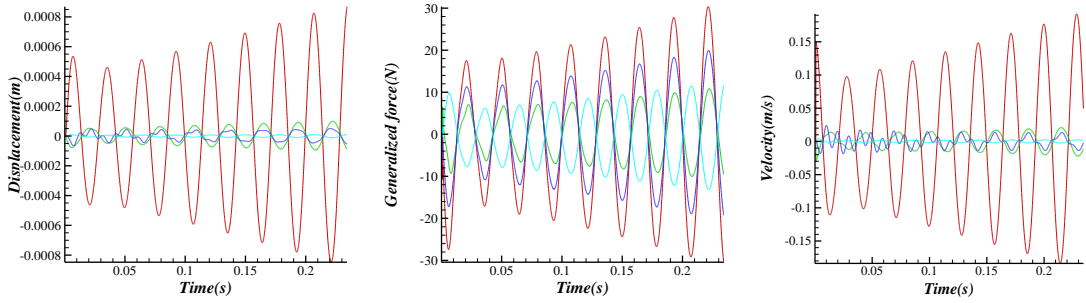

Fig A.4 $Ma$=0.83, $V$=100%$CFV_{ROM\text{-}AMS}$

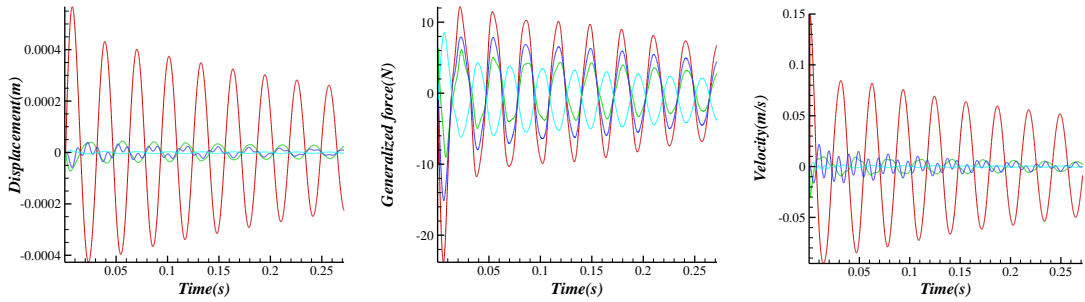

Fig A.5 $Ma$=0.87, $V$=95%$CFV_{ROM\text{-}AMS}$

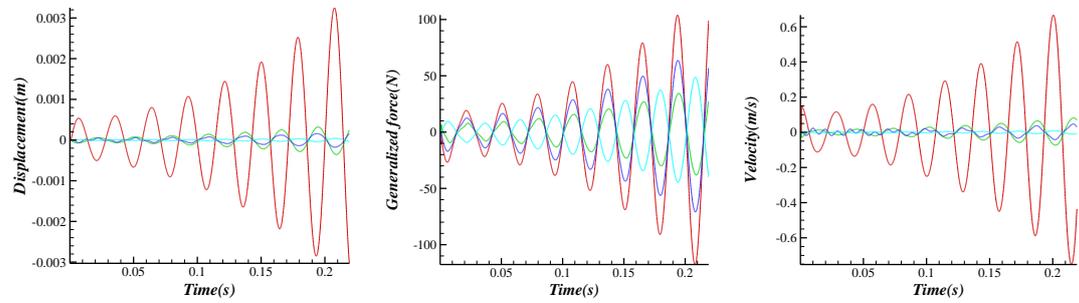

Fig A.6 $Ma$=0.87, $V$=100%$CFV_{ROM\text{-}AMS}$

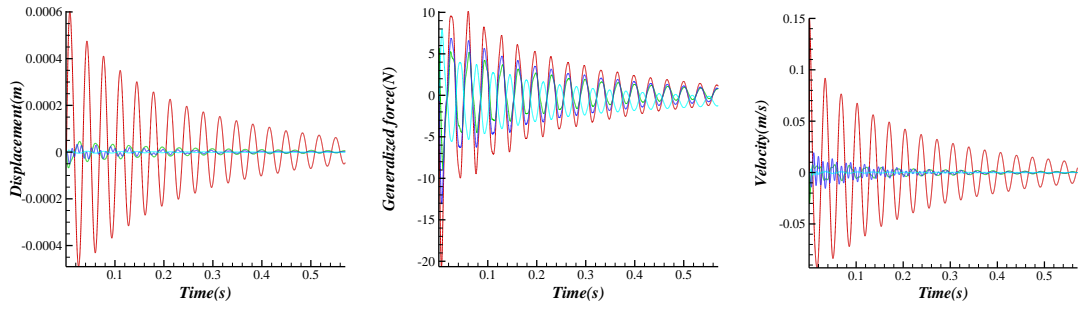

Fig A.7 *Ma*=0.91, *V*=90%*CFV$_{ROM-AMS}$*

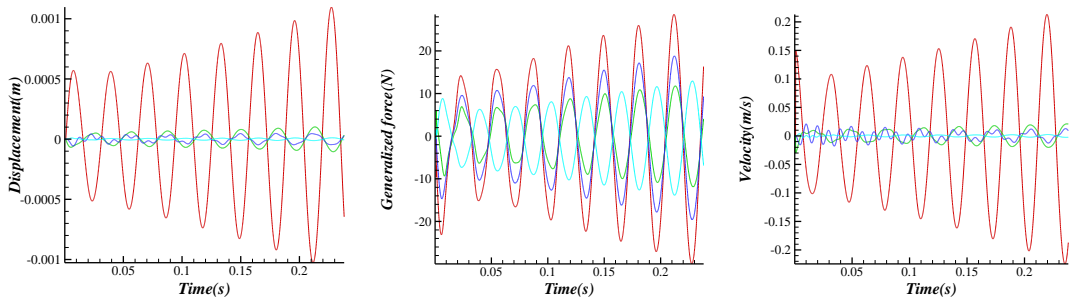

Fig A.8 *Ma*=0.91, *V*=95%*CFV$_{ROM-AMS}$*

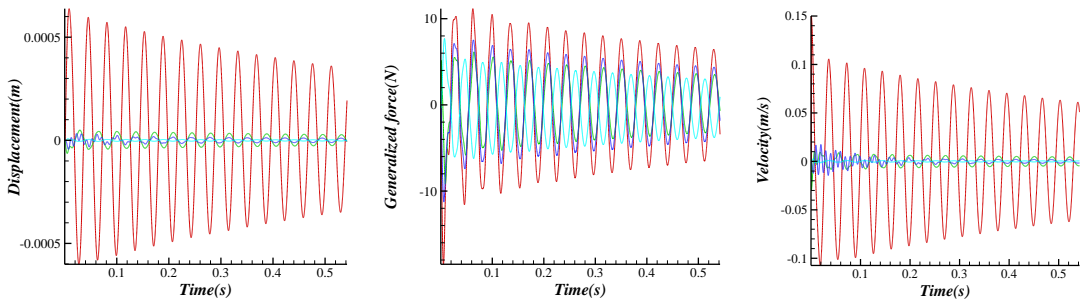

Fig A.9 *Ma*=0.95, *V*=87%*CFV$_{ROM-AMS}$*

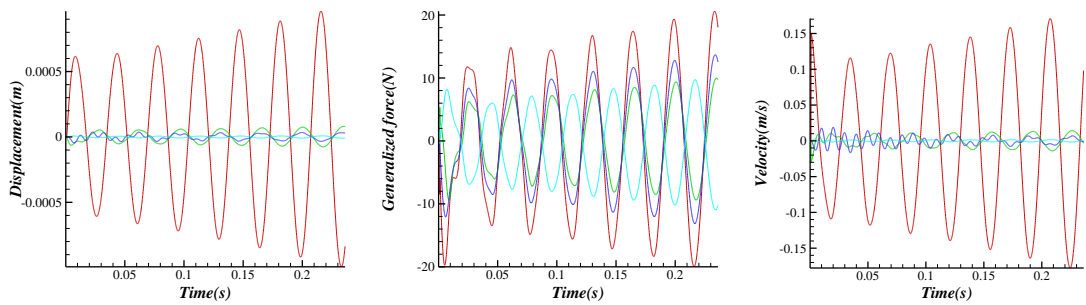

Fig A.10 *Ma*=0.95, *V*=90%*CFV$_{ROM-AMS}$*